\documentclass[aps,prb,showpacs,twocolumn,superscriptaddress]{revtex4} 

\usepackage{graphicx}
\usepackage[version=3]{mhchem}

\begin{document}

\title{Evolutionary construction of formation energy convex hull: Practical scheme and application to carbon-hydrogen binary system}



\author{Takahiro Ishikawa}%
 \email{ISHIKAWA.Takahiro@nims.go.jp}
 \affiliation{%
 ESICMM, National Institute for Materials Science, 1-2-1 Sengen, Tsukuba, Ibaraki 305-0047, Japan
 }%
\author{Takashi Miyake}%
 \affiliation{%
 ESICMM, National Institute for Materials Science, 1-2-1 Sengen, Tsukuba, Ibaraki 305-0047, Japan
 }%
 \affiliation{%
 CD-FMat, National Institute of Advanced Industrial Science and Technology, 1-1-1 Umezono, Tsukuba, Ibaraki 305-8568, Japan
 }%

\date{\today}

\begin{abstract}
We present an evolutionary construction technique of formation energy convex hull to search for thermodynamically stable compounds. In this technique, candidates with a wide variety of chemical compositions and crystal structures are created by systematically applying evolutionary operators, ``mating'', ``mutation'', and ``adaptive mutation'', to two target compounds, and the convex hull is directly updated through the evolution. 
We applied the technique to carbon-hydrogen binary system at 10\,GPa and obtained 15 hydrocarbons within the convex hull distance less than 0.5\,mRy/atom: graphane, polybutadiene, polyethylene, butane, ethane, methane, three molecular compounds of ethane and methane, and six molecular compounds of methane and hydrogen. These results suggest that our evolutionary construction technique is useful for the exploration of stable phases under extreme conditions and the synthesis of new compounds. 
\end{abstract}


\maketitle


\section{Introduction}
Search for thermodynamically stable compounds is crucial for the design and synthesis of novel functional materials and the understanding of the behavior of materials under extreme conditions. 
For example, an impressive achievement with respect to it is the discovery of high-temperature superconductivity in hydrides stabilized under high pressure conditions~\cite{Drozdov2015,Somayazulu2019,Drozdov2019-LaHx,Guigue2017,Goncharov2017,Geballe2018,Sun2019,Ishikawa2019}. 
In first-principles calculations, thermodynamically stable phases of materials are predicted by constructing a convex hull with respect to formation energy of the compounds. If a binary compound, $\alpha_{1-x}\beta_{x}$, is considered as an example, thermodynamically stable chemical compositions are typically explored through the following four steps: (1) determination of the most stable crystal structure at a fixed composition rate $x$ using structure search techniques such as random sampling~\cite{Pickard2011}, evolutionary algorithm~\cite{Oganov06}, particle-swarm optimization~\cite{Wang2010}, \textit{etc.}, (2) calculation of formation energy $E$ of the compound with respect to $\alpha$ and $\beta$, (3) repeat of the same calculations changing $x$, and (4) construction of the convex hull curve by plotting $E$ for $x$. The compositions on the convex hull correspond to thermodynamically stable ones. This approach has been successful in the discovery of new materials including superconducting hydrides mentioned above, whereas it requires greater computational resources, \textit{i.e.} a large number of structural optimizations, to construct the convex hull accurately. 
Moreover, many calculations end up in a wasted effort because the appearance of the compounds on the convex hull is a rare event. 
Therefore, it is important for the increase of the search efficiency to 
use a direct construction approach of the convex hull without the procedure through the four steps, and various improvements have been carried out~\cite{Oganov2010-book,Oganov2010-variablecomposition}. 

In this study, we propose an approach for the direct search, 
which we call ``evolutionary construction technique of formation energy convex hull''. 
In this technique, candidates with a wide variety of chemical compositions and crystal structures are created by systematically applying evolutionary operators, ``mating'', ``mutation'' consisting of permutation, distortion, reflection, modulation, addition, elimination, and substitution, and ``adaptive mutation'', to stable and metastable compounds on or near the convex hull. After the structural optimizations, the convex hull is updated. By repeatedly performing this process, the convex hull is evolutionarily constructed, which achieves more efficient search for thermodynamically stable compounds compared with the typical approach based on the exhaustive search. In this paper, we show the details of the method and its application to carbon-hydrogen binary system. 

\section{Details of evolutionary construction technique}
\begin{figure}
\includegraphics[width=8.2cm]{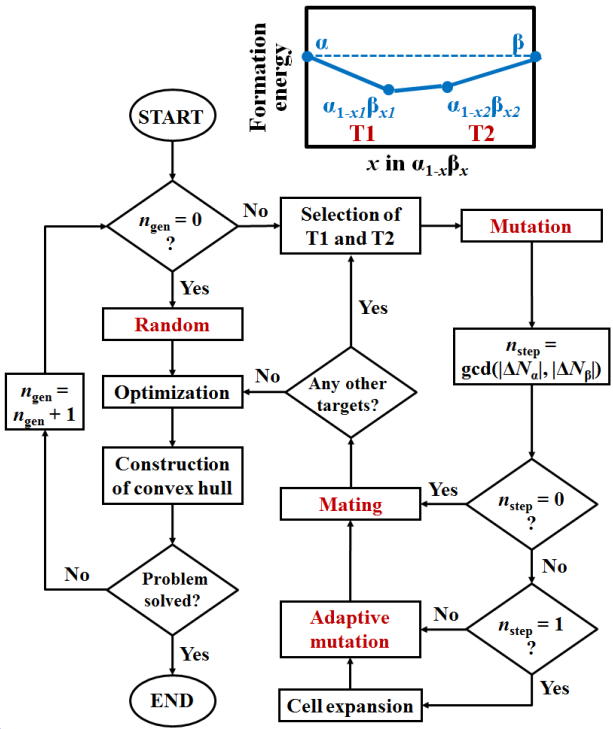}
\caption{\label{Fig-flowchart} 
(Color online) Flowchart for the evolutionary construction of formation energy convex hull.}
\end{figure}
\begin{figure}
\includegraphics[width=8.2cm]{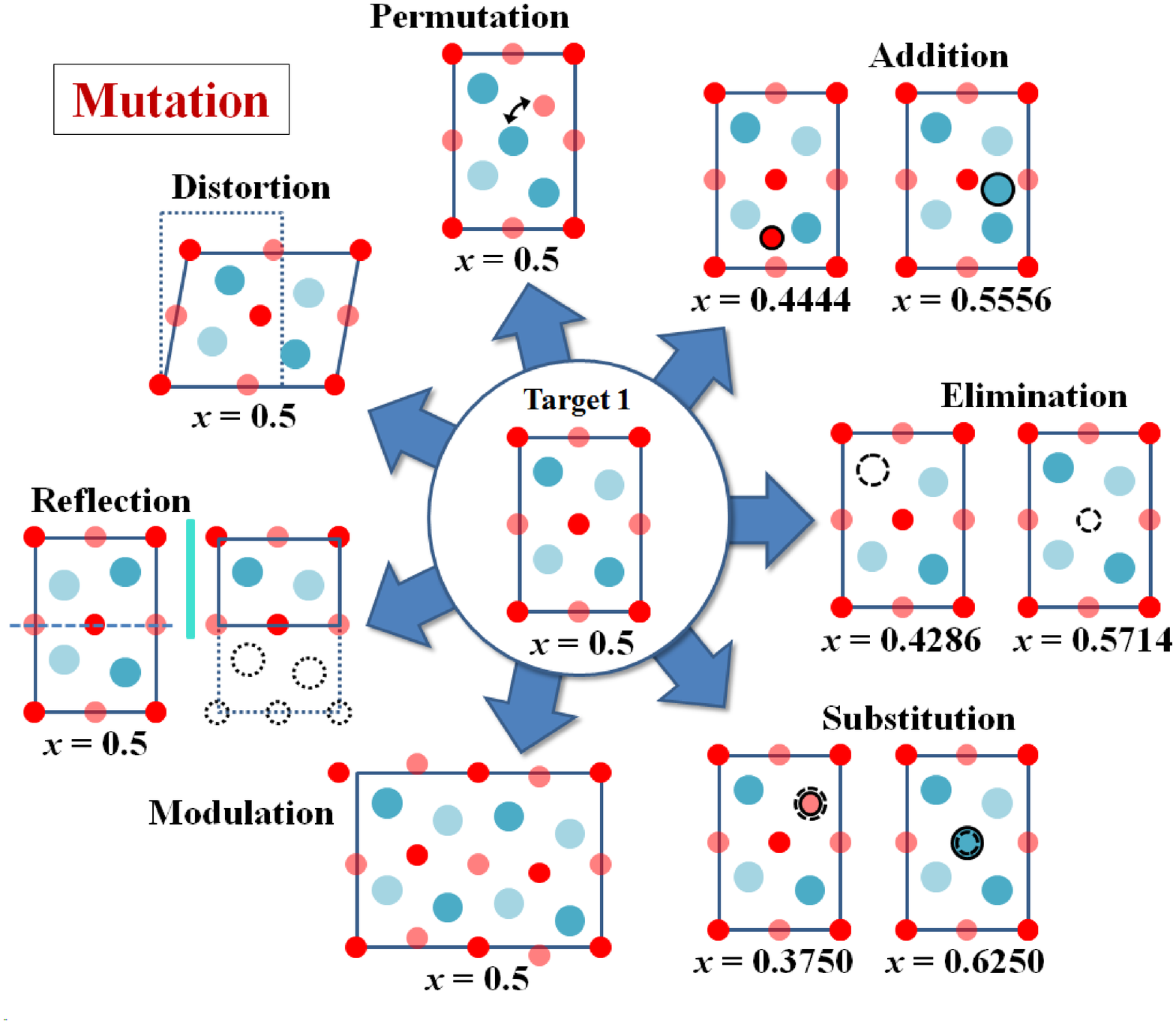}
\caption{\label{Fig-mutation} 
(Color online) Application of the mutation to a target compound, which consists of seven operators, permutation, distortion, reflection, modulation, addition, elimination, and substitution. The composition $x$ is defined as $N_{\text{l}} / (N_{\text{s}} + N_{\text{l}})$, where $N_{\text{s}}$ ($N_{\text{l}}$) is the number of the small (large) balls included in the cell.}
\end{figure}
\begin{figure}
\includegraphics[width=8.2cm]{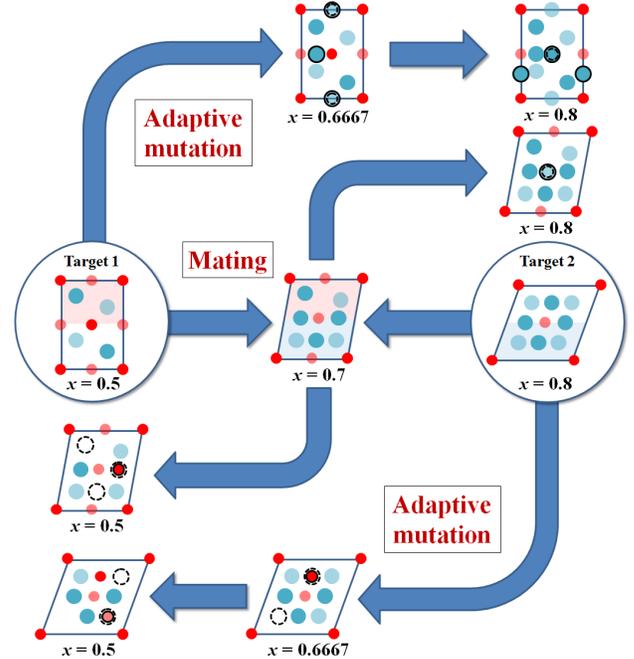}
\caption{\label{Fig-nstep2} 
(Color online) Application of the adaptive mutation and the mating to two target compounds. The composition $x$ is defined as $N_{\text{l}} / (N_{\text{s}} + N_{\text{l}})$, where $N_{\text{s}}$ ($N_{\text{l}}$) is the number of the small (large) balls included in the cell. The added, eliminated, and substituted atoms are represented as the solid, broken, and solid-broken circles, respectively.}
\end{figure}
Evolutionary algorithm requires setting and tunning many parameters such as 
``heredity rate'', ``permutation rate'', ``mutation rate'', ``selection rate'', ``population size'', and so on.
On the other hand, in our technique the parameter setting and tuning 
are reduced because potential candidates are systematically created according to rules described below. 
Figure \ref{Fig-flowchart} shows the flowchart of the evolutionary construction technique. 
Let us consider the case of a binary system, $\alpha$ ($x = 0$) and $\beta$ ($x = 1$).   
At the preliminary generation ($n_{\text{gen}} = 0$), chemical compositions and crystal structures are generated by randomly mixing the two elements, and the structural optimizations are performed. Then, the formation energies from the elements are calculated, and the preliminary convex hull is constructed. 

From the first generation ($n_{\text{gen}} = 1$), two target compounds, T1 and T2, are freely selected from the compounds which emerge on or near the convex hull. Then, the evolutionary operator, mutation, is applied to each of T1 and T2, as shown in Fig. \ref{Fig-mutation}. The mutation consists of seven operators as follows: permutation, distortion, reflection, modulation, addition, elimination, and substitution. 
The permutation is the operator that the positions are exchanged between different atomic species. The distortion is the operator that the lattice parameters, \textit{i.e.} three lengths and three angles, are randomly changed. 
The reflection is the operator that atomic positions are partially converted according to a reflection about a mirror on the $ab$, $bc$, or $ca$ plane. 
The modulation is the operator that a zigzag modulation is randomly provided to 
the calculation cell doubled along the axis with the smallest length of the three, which is effective for the discovery of long-period structures such as modulated structures observed in elemental metals under high pressure~\cite{McMahon06,Ishikawa06,Ishikawa08}. 
In contrast to the above four operators, the addition, elimination, and substitution are the operators varying the chemical composition $x$. 
The addition is the operator that the atom is randomly added in the calculation cell. 
The elimination is the operator that crystal structures with a wide variety of $x$ are created by step-by-step eliminating the atoms in the cell. Similarly, the substitution is the operator that the structures are created by step-by-step substituting the atomic species. For example, in the case of Fig. \ref{Fig-mutation}, $x$ is gradually decreased (increased) to 0 (1) by eliminating the large (small) balls or substituting the large (small) ones for the small (large) ones. 
Especially, these operators are effective for the exploration with respect to the case that the structure is basically similar, although the chemical composition is different, \textit{e.g.} low-, intermediate, and high-$T_{\text{c}}$ superconducting phases predicted in compressed sulfur hydrides~\cite{Ishikawa2016-SciRep,Akashi2016}. 

\begin{table}
\caption{\label{mutation-case}
Operations for the atoms in the calculation cell with respect to the adaptive mutation.}
\begin{ruledtabular}
\begin{tabular}{lcc}
    Case & & Operation\\ \hline
    (i) $\delta N_{\alpha} \geq 0 \cap \delta N_{\beta} \geq 0$ &  & addition\\
    (ii) $\delta N_{\alpha} \leq 0 \cap \delta N_{\beta} \leq 0$ &  & elimination\\
    (iii) others &(a) $\delta N_{\alpha} + \delta N_{\beta} = 0$  & substitution\\
     &(b) $\delta N_{\alpha} + \delta N_{\beta} > 0$  & sub. + add.\\
     &(c) $\delta N_{\alpha} + \delta N_{\beta} < 0$  & sub. + elm.\\
\end{tabular}
\end{ruledtabular}
\end{table}
Next, the adaptive mutation and the mating are performed, in which the candidates are created using the structural information about both T1 and T2 (Fig. \ref{Fig-nstep2}). 
The adaptive mutation is a controversial theory of biological evolution~\cite{Cairns1988,Hendrickson2002} and a hypothesis that the mutations are much less random and more purposeful than those considered in traditional evolutionary theory. 
In this study we adopted this idea in crystal structure search algorithm. The adaptive mutation is the operator that the structures are created by step-by-step transforming T1 (T2) until the composition $x$ and the number of the formula unit included in the calculation cell, $n_{\text{f.u.}}$, coincide with those of T2 (T1). 
Here, we represent variations of the number of the $\alpha$ and $\beta$ atoms according to the transformation as 
$\Delta N_{\alpha}$ and $\Delta N_{\beta}$, respectively. 
The number of the adaptive mutation steps, $n_{\text{step}}$, is defined as the greatest common divisor (gcd) of the absolute values of $\Delta N_{\alpha}$ and $\Delta N_{\beta}$, \textit{i.e.} $n_{\text{step}} = \text{gcd}(|\Delta N_{\alpha}|, |\Delta N_{\beta}|)$. 
The number of the atoms in the calculation cell is varied by $\delta N_{\alpha} = \Delta N_{\alpha} / n_{\text{step}}$ for $\alpha$ and $\delta N_{\beta} = \Delta N_{\beta} / n_{\text{step}}$ for $\beta$ at each step. 
The atoms in the unit cell are operated according to five cases determined by the condition of $\delta N_{\alpha}$ and $\delta N_{\beta}$, as shown in Table \ref{mutation-case}. 
For example, in Fig. \ref{Fig-nstep2}, $\delta N_{\alpha}$ (small ball) and $\delta N_{\beta}$ (large ball) are -1 and 2 for the transformation from T1 to T2, respectively. The value of $\delta N_{\alpha} + \delta N_{\beta}$ 
is equal to 1, which is classified into the case (iii)-(b); first a large ball is substituted for a small ball selected randomly, and then a large ball is randomly added. 
At this stage, the composition is increased to $x = 0.6667$, which is intermediate one between T1 and T2. Performing the similar procedure again, a candidate structure with $x = 0.8$ is created, based on the structural information of T1. 
In contrast, when the transformation starts from T2, corresponding to the case (iii)-(c), 
candidate structures with $x=0.6667$ and 0.5 are created by the random elimination of the large balls in addition to the substitution, based on the structural information of T2. 

The mating is the operator that a structure is created by combining T1 and T2, which is similar to the ``heredity'' operator proposed by Oganov \textit{et al.}~\cite{Oganov06}. First, each structure is cut into two at the plane, 
which passes through a point $s$ ($0 \leq s < 1$) on the axis randomly selected and is parallel to the plane formed by the other two axes. Then, 
the $[0, s)$ region for T1 and the $[s, 1)$ region for T2 are combined. The calculation cell of the combined structure is obtained by mixing the lattice parameters of T1 and those of T2 with a rate of $s:1-s$. In this way, a candidate structure is created with a composition different from T1 and T2, and further two structures are created from the structure, as with the case of the adaptive mutation (Fig. \ref{Fig-nstep2}). 

Note that there are three cases with respect to $n_{\text{step}}$: $n_{\text{step}} \geq 2$, $n_{\text{step}} = 1$, and $n_{\text{step}} = 0$ (Fig. \ref{Fig-flowchart}). 
For the first case, both the adaptive mutation and the mating are applied  as shown in Fig. \ref{Fig-nstep2}. 
For the second case, cell expansion is performed before the application of the adaptive mutation and mating because no intermediate chemical compositions between T1 and T2 are created by the procedure mentioned above. First the calculation cell is doubled along the axis with the smallest length of the three for both T1 and T2, and then the adaptive mutation and mating are applied to the expanded cells. 
This procedure achieves the increase of $n_{\text{step}}$ to 2 and the creation of the intermediate compounds between T1 and T2. For the third case, only the mating is applied because T1 has the same $x$ and $n_{\text{f.u.}}$ as T2 and the adaptive mutation is unnecessary. If computational resources are still available, further structures can be created using other T1 and T2. Then, the structural optimizations are performed and the convex hull is updated. 

\section{Application to carbon-hydrogen binary system}
\begin{figure}
\includegraphics[width=8.0cm]{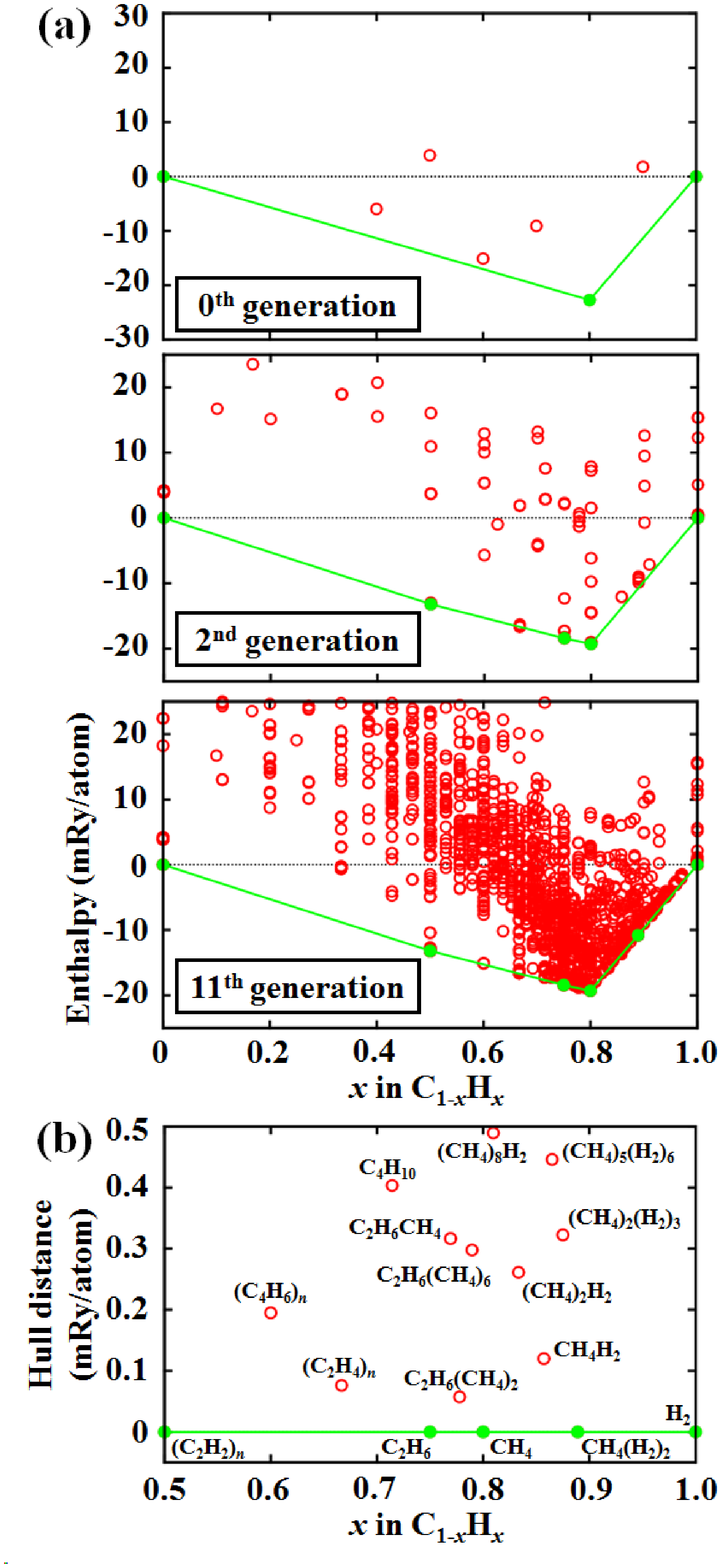}
\caption{\label{Fig-CHEvolution} 
(Color online) (a) Evolution of formation energy convex hull with respect to \ce{C_{1-$x$}H_{$x$}} at 10\,GPa. (b) Compounds with the convex hull distance less than 0.5\,mRy/atom.}
\end{figure}
\begin{table}
\caption{\label{results}
Compounds with the hull distance $\Delta H$ less than 0.5\,mRy/atom.}
\begin{ruledtabular}
\begin{tabular}{ccccc}
$x$ &  & & $\Delta H$ & Band gap\\
  &  & & (mRy/atom) & (eV)\\ \hline
0 & \ce{C} & diamond & 0 & 4.3\\
0.5 & \ce{(C_{2}H_{2})_{$n$}} & graphane & 0 & 4.7\\
0.6 & \ce{(C_{4}H_{6})_{$n$}} & polybutadiene& 0.19 & 6.1\\
0.6667 & \ce{(C_{2}H_{4})_{$n$}} & polyethylene & 0.08 & 6.2\\
0.7143 & \ce{C_{4}H_{10}} & butane & 0.40& 6.7\\
0.75 & \ce{C_{2}H_{6}} & ethane & 0 & 7.7\\
0.7692 & \ce{C_{2}H_{6}CH_{4}} & eth. + met. & 0.32& 8.0\\
0.7778 & \ce{C_{2}H_{6}(CH_{4})_{2}} & eth. + met. & 0.06& 7.9\\
0.7895 & \ce{C_{2}H_{6}(CH_{4})_{6}} & eth. + met. & 0.30& 8.1\\
0.8 & \ce{CH_{4}} & methane & 0 & 8.3\\
0.8095 & \ce{(CH_{4})_{8}H_{2}} & met. + hyd.& 0.49& 8.0\\
0.8333 & \ce{(CH_{4})_{2}H_{2}} & met. + hyd.& 0.26& 8.3\\
0.8571 & \ce{CH_{4}H_{2}} & met. + hyd.& 0.12& 8.2\\
0.8649 & \ce{(CH_{4})_{5}(H_{2})_{6}} & met. + hyd.& 0.45& 8.4\\
0.875 & \ce{(CH_{4})_{2}(H_{2})_{3}} & met. + hyd.& 0.32& 7.8\\
0.8889 & \ce{CH_{4}(H_{2})_2} & met. + hyd.& 0 & 7.9\\
1 & \ce{H_{2}} & hydrogen & 0 & 7.3\\
\end{tabular}
\end{ruledtabular}
\end{table}
\begin{figure*}
\includegraphics[width=15.0cm]{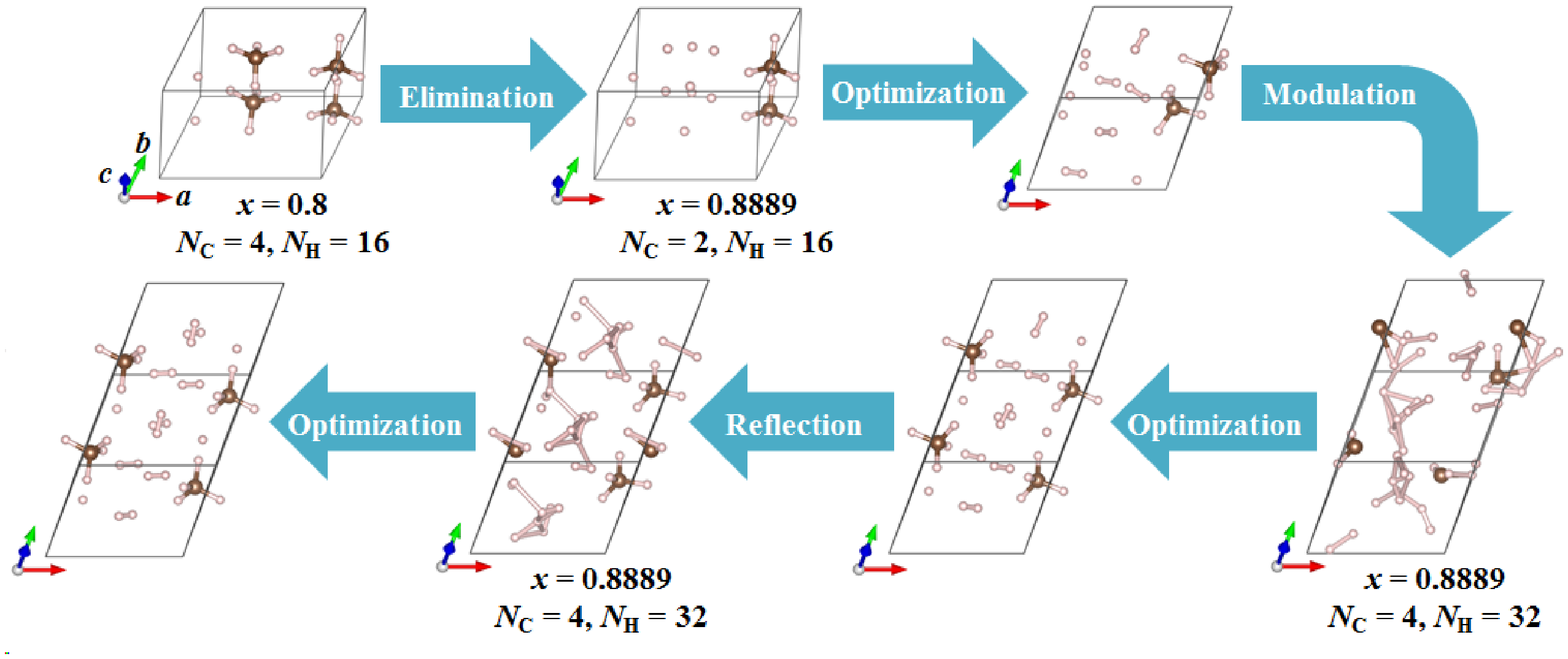}
\caption{\label{Fig-transformation} 
(Color online) Formation process of the compound with $x = 0.8889$ obtained through the evolutionary construction technique. $N_{\text{C}}$ and $N_{\text{H}}$ represent the number of carbon atoms in the calculation cell and that of hydrogen atoms, respectively. At the sixth generation, the two carbon atoms in the compound with $x = 0.8$ are eliminated from the cell by the evolutionary operator ``elimination'', and four \ce{H_{2}} molecules are formed by the structural optimization. At the seventh generation, $N_{\text{C}}$ and $N_{\text{H}}$ are doubled and the atoms are displaced by the ``modulation'' along the $b$ axis, and four \ce{CH_{4}} molecules and eight \ce{H_{2}} molecules are moved to more stable positions by the optimization. 
At the tenth generation, the positions of the atoms included in the region of $0 \leq a < 0.5$ are converted by ``reflection'' about a mirror on the $ab$ plane, and then 
the most stable structure is obtained by the optimization, in which the hydrogen atoms are slightly displaced from the positions before the reflection. Crystal structures were drawn with VESTA~\cite{VESTA}.}
\end{figure*}
\begin{figure*}
\includegraphics[width=16.0cm]{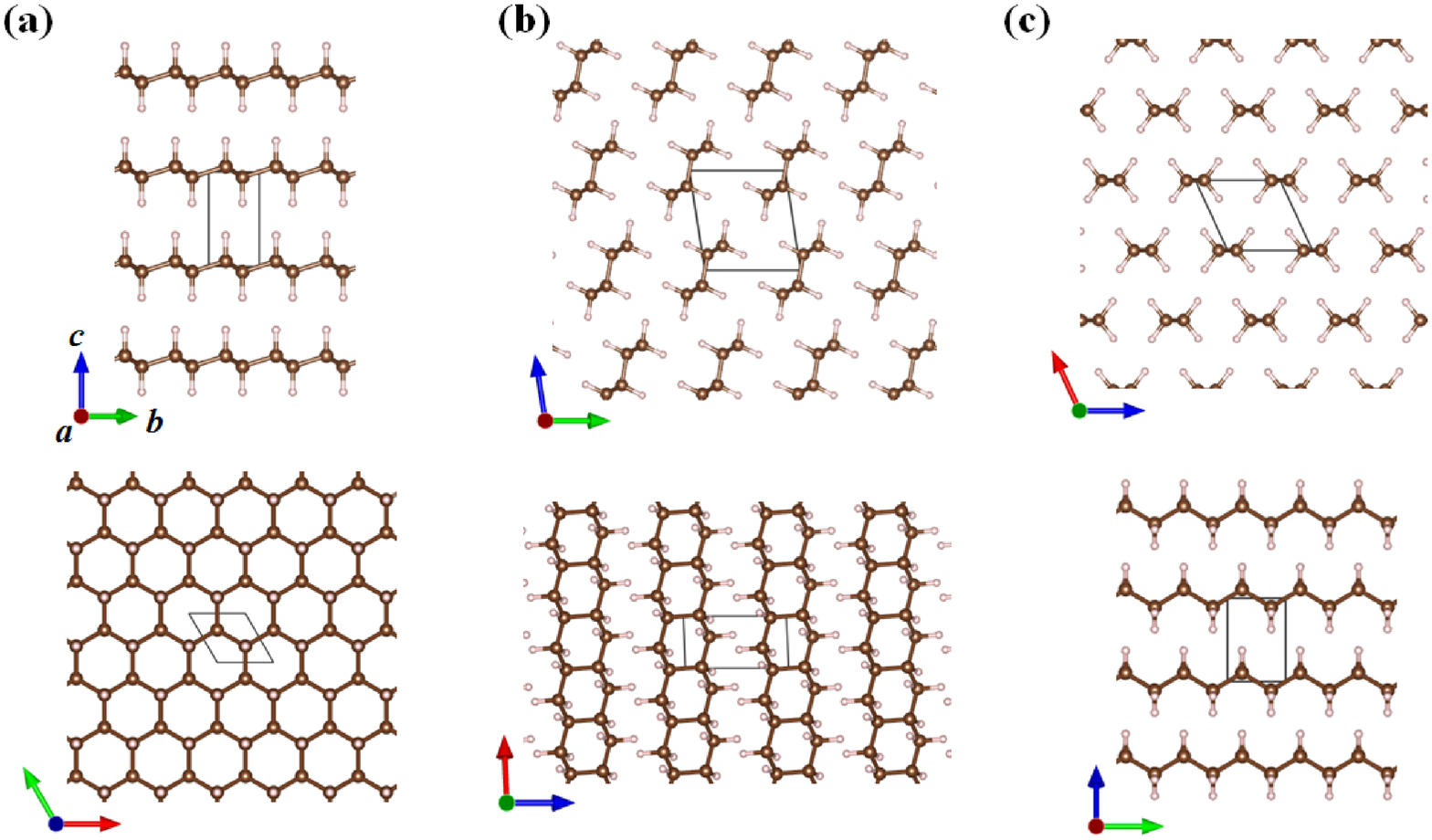}
\caption{\label{Fig-structure-type1} 
(Color online) Crystal structures at 10\,GPa: (a) \ce{(C_{2}H_{2})_{$n$}}, (b) \ce{(C_{4}H_{6})_{$n$}}, and (c) \ce{(C_{2}H_{4})_{$n$}}. Large and small balls represent carbon and hydrogen atoms, respectively. Crystal structures were drawn with VESTA~\cite{VESTA}.}
\end{figure*}
\begin{table}
\caption{\label{structureparameter-type1}
Structural parameters of \ce{(C_{4}H_{6})_{$n$}} and \ce{(C_{2}H_{4})_{$n$}}.}
\begin{ruledtabular}
\begin{tabular}{llll}
& & Cell (\AA, $^{\circ}$) & Atomic position \\ \hline
\ce{(C_{4}H_{6})_{$n$}} &$C2/m$ & $a$ 9.2041 & C $4i$ 0.5266 0 0.1569 \\
& & $b$ 2.5128 & C $4i$ 0.8869 0 0.7557 \\
& & $c$ 4.8316 & H $4i$ 0.4284 0 0.2612\\
& & $\beta$ 98.70 & H $4i$ 0.1485 0 0.4707\\
& &  & H $4i$ 0.7858 0 0.8510 \\ \hline
\ce{(C_{2}H_{4})_{$n$}} & $P2_{1}/m$ & $a$ 3.6115 & C $2e$ 0.4983 0.25 0.3946 \\
&  & $b$ 2.5210 & H $2e$ 0.7594 0.25 0.3224\\
&  & $c$ 3.9636 & H $2e$ 0.2296 0.25 0.1272\\
&  & $\beta$ 114.82 & \\
\end{tabular}
\end{ruledtabular}
\end{table}
\begin{figure*}
\includegraphics[width=16.0cm]{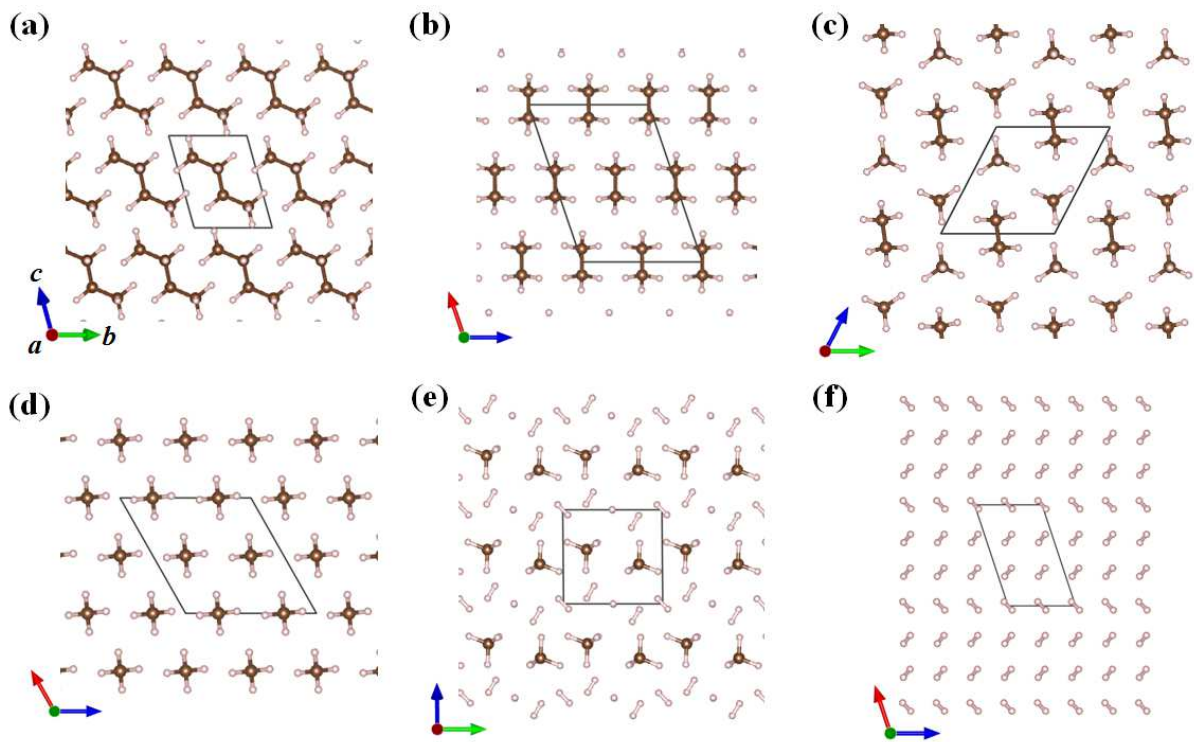}
\caption{\label{Fig-structure-type2} 
(Color online) Crystal structures at 10\,GPa: (a) \ce{C_{4}H_{10}}, (b) \ce{C_{2}H_{6}}, (c) \ce{C_{2}H_{6}(CH_{4})_{2}}, (d) \ce{CH_{4}}, (e) \ce{CH_{4}(H_{2})_{2}}, and (f) \ce{H_{2}}. Large and small balls represent carbon and hydrogen atoms, respectively. Crystal structures were drawn with VESTA~\cite{VESTA}.}
\end{figure*}
\begin{table}
\caption{\label{structureparameter-type2}
Structural parameters of \ce{C_{4}H_{10}}, \ce{C_{2}H_{6}}, \ce{C_{2}H_{6}(CH_{4})_{2}}, \ce{CH_{4}}, \ce{CH_{4}(H_{2})_{2}}, and \ce{H_{2}}.}
\begin{ruledtabular}
\begin{tabular}{llll}
& & Cell (\AA, $^{\circ}$) & Atomic position \\ \hline
\ce{C_{4}H_{10}} & $P$-1 & $a$ 3.7519 & C $2i$ 0.9532 0.2335 0.7636 \\
&  & $b$ 4.5375 & C $2i$ 0.1387 0.5177 0.6456 \\
&  & $c$ 5.3520 & H $2i$ 0.2062 0.7489 0.7741 \\
&  & $\alpha$ 94.70 & H $2i$ 0.6742 0.2247 0.7914 \\
&  & $\beta$ 109.37 & H $2i$ 0.1777 0.2493 0.9618 \\
&  & $\gamma$ 112.77 & H $2i$ 0.1432 0.0011 0.3711 \\
&  &  & H $2i$ 0.5560 0.4574 0.3470 \\ \hline
\ce{C_{2}H_{6}} & $C2/c$ & $a$ 8.6551 & C $8f$ 0.5835 0.0919 0.0364 \\
&  & $b$ 3.3924 & H $8f$ 0.6038 0.2681 0.9023 \\
&  & $c$ 6.2579 & H $8f$ 0.1788 0.3679 0.0847 \\
&  & $\beta$ 108.70 & H $8f$ 0.8990 0.2138 0.8189 \\ \hline
\ce{C_{2}H_{6}(CH_{4})_{2}} & $P$-1 & $a$ 3.4521 & C $2i$ 0.0795 0.4212 0.1286 \\
&  & $b$ 5.6046 & C $2i$ 0.3595 0.1591 0.6933 \\
&  & $c$ 5.8933 & H $2i$ 0.2635 0.5498 0.1716 \\
&  & $\alpha$ 61.74 & H $2i$ 0.7505 0.6406 0.3646 \\
&  & $\beta$ 85.06 & H $2i$ 0.3245 0.8289 0.3165 \\
&  & $\gamma$ 89.97 & H $2i$ 0.8449 0.3304 0.2843 \\
&  &  & H $2i$ 0.7569 0.9805 0.1095 \\
&  &  & H $2i$ 0.2560 0.2577 0.1337 \\
&  &  & H $2i$ 0.2733 0.0858 0.5636 \\ \hline
\ce{CH_{4}} & $C2/c$ & $a$ 6.2809 & C $4e$ 0 0.0964 0.25 \\
&  & $b$ 3.3144 & H $8f$ 0.8386 0.0937 0.6830 \\
&  & $c$ 6.1895 & H $8f$ 0.4831 0.2144 0.5992 \\
&  & $\beta$ 119.59 & \\ \hline
\ce{CH_{4}(H_{2})_2} & $P$-1 & $a$ 3.7932 & C $2i$ 0.7066 0.7632 0.4235 \\
&  & $b$ 5.1936 & H $2i$ 0.5864 0.3091 0.1891 \\
&  & $c$ 5.2688 & H $2i$ 0.6763 0.9573 0.3438 \\
&  & $\alpha$ 90.46 & H $2i$ 0.1761 0.2273 0.3559 \\
&  & $\beta$ 111.02 & H $2i$ 0.1048 0.3459 0.6505 \\
&  & $\gamma$ 90.09 & H $2i$ 0.9818 0.0495 0.9461 \\
&  &  & H $2i$ 0.5690 0.3301 0.6422 \\
&  &  & H $2i$ 0.9006 0.5007 0.9977 \\
&  &  & H $2i$ 0.4611 0.7652 0.9402 \\ \hline
\ce{H_{2}} & $P2_{1}/c$ & $a$ 5.4733 & H $4e$ 0.9589 0.0649 0.5394 \\
&  & $b$ 3.4470 & H $4e$ 0.7078 0.5718 0.4044 \\
&  & $c$ 3.4760 & H $4e$ 0.3733 0.5759 0.7347 \\
&  & $\beta$ 108.39 & \\
\end{tabular}
\end{ruledtabular}
\end{table}

We applied the evolutionary construction technique to carbon-hydrogen binary system (\ce{C_{1-$x$}H_{$x$}}) under high pressure. 
Carbon, hydrogen, and hydrocarbon have profoundly affected humankind, and  
the knowledge of thermodynamically stable phases of \ce{C_{1-$x$}H_{$x$}} system is important for materials science, earth and planetary science, life science, and so on. 
Gao \textit{et al.} predicted that methane, \ce{CH_{4}} ($x = 0.8$), dissociates into ethane, \ce{C_{2}H_{6}} ($x = 0.75$), at 95\,GPa, 
butane, \ce{C_{4}H_{10}} ($x = 0.7143$), at 158\,GPa, and 
diamond ($x = 0$) at 287\,GPa at zero temperature~\cite{Gao2010}. 
Liu \textit{et al.} searched for stable phases of the binary system in pressure range of 
100-300\,GPa and predicted that ethylene, \ce{C_{2}H_{4}} ($x = 0.6667$), \ce{C_{2}H_{6}}, and  \ce{CH_{4}} emerge on the convex hull at 100\,GPa, and \ce{CH_{4}} and \ce{C_{2}H_{6}} become unstable at 100\,GPa and 200\,GPa, respectively~\cite{Liu2016}. 
These results indicate that hydrogen-rich hydrocarbons get to be unstable with the increase of pressure. In contrast, there is a possibility that 
novel hydrocarbons are stabilized in low pressure region, and 
we explored them by performing the evolutionary construction technique at the pressure of 10\,GPa.   

First, we developed the calculation code of the evolutionary construction technique following the flowchart shown in Fig. \ref{Fig-flowchart}, and combined it with the Quantum ESPRESSO (QE) code~\cite{QE} to perform the optimization for the structures created by the evolutionary operators. 
In this study, we intentionally used no experimental and theoretical results on the stable compositions and structures reported earlier in order to verify the prediction ability of our search technique. 
For the preliminary generation ($n_{\text{gen}} = 0$), we prepared for the calculation cell including 10 carbon 
atoms and step-by-step substituted a hydrogen atom for a carbon one. In this way, $x$ was increased from 0 to 1 with an interval of 0.1, and 11 compositions were created. The lattice parameters and atomic positions were randomly generated at each $x$. 
The number of the atoms in the calculation cell, which is increased by the addition, the mating, the modulation, and the cell expansion, was limited to less than 80. 
The atoms were displaced by $\pm 0.05$ in fractional coordinates with respect to the modulation. 
The pressure was set at 10\,GPa. 
We used Perdew, Burke and Ernzerhof~\cite{PBE} for the exchange-correlation functional, and 
the Rabe-Rappe-Kaxiras-Joannopoulos ultrasoft pseudopotential~\cite{RRKJ90}. 
The $k$-space integration over the Brillouin zone (BZ) was performed 
on a 16 $\times$ 16 $\times$ 16 grid for the compounds with 1-4 atoms in the calculation cell, 
12 $\times$ 12 $\times$ 12 for 5-8 atoms, and 
8 $\times$ 8 $\times$ 8 for more than 8 atoms. 
The energy cutoffs were set at 100\,Ry for wave function and 800\,Ry for charge density, respectively.  

Figure \ref{Fig-CHEvolution} (a) shows the evolution of the formation enthalpy convex hull with respect to the composition $x$ in \ce{C_{1-$x$}H_{$x$}}. 
The convex hull is roughly converged at the second generation, in which the compounds with $x=0.5$, 0.75, and 0.8 appear on the convex hull, and the compound with $x=0.8889$ emerges on the convex hull at the sixth generation. 
The compound with $x=0.8889$ is first created by applying the operator ``elimination'' to the compound with $x = 0.8$ including four \ce{CH_{4}} molecules in the calculation cell 
 and then is refined by applying ``modulation'' at the seventh generation and ``reflection'' at the tenth generation (Fig. \ref{Fig-transformation}). 
The result at the 11th generation shows that the hull distance is gradually decreased with the increase of $x$ and many hydrogen-rich compounds are close to the convex hull, which implies that hydrogen-rich compounds are more stable than carbon-rich ones at 10 GPa. 

Figure \ref{Fig-CHEvolution} (b) shows the compounds included in the region that the distance from the convex hull ($\Delta H$) is less than 0.5\,mRy/atom. 
In Ref. \cite{Wu2013}, the authors reported that more than 80\% of experimentally synthesized compounds included in the Inorganic Crystal Structure Database (ICSD) have $\Delta H$ less than 36 meV/atom ($= 2.65$ mRy/atom). This error is associated with the approximations used in first-principles calculations and the omission of temperature effects. Therefore, we used the tolerance of $\Delta H < 0.5$\,mRy/atom, which is less than one-fifth of the error, to search for potential compounds that can be synthesized by experiments. All the compounds with $\Delta H < 0.5$\,mRy/atom, \textit{i.e.} 15 hydrocarbons, are listed in 
Table \ref{results}. 
The compounds with $x = 0.5$, 0.6, and 0.6667 are formed by polymerization of  \ce{C_{2}H_{2}}, \ce{C_{4}H_{6}}, and \ce{C_{2}H_{4}} molecules, respectively. 
Here, we 
indicate the polymerization with the subscript $n$. 
$(\ce{C_{2}H_{2}})_{n}$ takes a trigonal $P$-3$m$1 structure polymerized two-dimensionally in the $ab$ plane (Fig. \ref{Fig-structure-type1} (a)), which has been known as the chair-type graphane~\cite{Sofo2007}. 
$(\ce{C_{4}H_{6}})_{n}$ (polybutadiene) and $(\ce{C_{2}H_{4}})_{n}$ (polyethylene) take a monoclinic $C2/m$ polymerized along the $a$ axis (Fig. \ref{Fig-structure-type1} (b)) and a monoclinic $P2_{1}/m$ polymerized along the $b$ axis (Fig. \ref{Fig-structure-type1} (c)), respectively. The structural parameters are listed in Table \ref{structureparameter-type1}. The compounds with $0.7143 \leq x \leq 1$ are formed by the condensation of isolated $\ce{C_{4}H_{10}}$, $\ce{C_{2}H_{6}}$, $\ce{CH_{4}}$, and $\ce{H_{2}}$ molecules. 
The structural parameters are listed in Table \ref{structureparameter-type2}. 
$\ce{C_{4}H_{10}}$ and $\ce{C_{2}H_{6}}$ take a triclinic $P$-1 (Fig. \ref{Fig-structure-type2} (a)) and a monoclinic $C2/c$ (Fig. \ref{Fig-structure-type2} (b)), respectively. 
$C2/c$ $\ce{C_{2}H_{6}}$ is energetically nearly degenerate with $P2_{1}/c$ $\ce{C_{2}H_{6}}$ predicted earlier~\cite{Gao2010}. 
In the region of $0.75 < x < 0.8$, molecular compounds of \ce{C_{2}H_{6}} and \ce{CH_{4}} are thermodynamically stable, and especially the compound with $\ce{C_{2}H_{6}}:\ce{CH_{4}} = 1:2$, \textit{i.e.} \ce{C_{2}H_{6}(CH_{4})_{2}}, with a triclinic $P$-1 (Fig. \ref{Fig-structure-type2} (c)) is energetically competing with the dissociated state, $\ce{C_{2}H_{6}} + \ce{CH_{4}}$: $\Delta H = 0.06$\,mRy/atom. 
\ce{CH_{4}} with $x=0.8$ takes a monoclinic $C2/c$ structure (Fig. \ref{Fig-structure-type2} (d)), which is energetically nearly degenerate with monoclinic $P2_{1}2_{1}2_{1}$ structure predicted earlier~\cite{Gao2010}. 
Similarly, in the region of $0.8 < x < 1$, molecular compounds of \ce{CH_{4}} and \ce{H_{2}} (hydrogen) are stabilized, and especially \ce{CH_{4}(H_{2})_{2}} with $\ce{CH_{4}}:\ce{H_{2}} = 1:2$, which takes a triclinic $P$-1 (Fig. \ref{Fig-structure-type2} (e)), emerges on the convex hull. 
For \ce{H_{2}} with $x=1$, we obtained a new structure with a monoclinic $P2_{1}/c$ (Fig. \ref{Fig-structure-type2} (f)), in which the orientation of the \ce{H_{2}} molecules is different from those of the structures predicted earlier~\cite{Pickard07}. 
The enthalpy of the $P2_{1}/c$ structure is lower by 0.03\,mRy/atom than that of a hexagonal $P6_{3}/m$ structure~\cite{Pickard07}, which are energetically nearly 
degenerate with each other. 

Our results suggest that the most hydrogen-rich hydrocarbon on the convex hull is \ce{CH_{4}(H_{2})_{2}} ($x = 0.8889$), which is qualitatively consistent with the results of earlier experiments~\cite{Somayazulu1996} and recent first-principles calculations~\cite{Conway2019}. 
The compounds with $x > 0.8889$ is thermodynamically unstable, and $\Delta H$ is increased with increase of $x$. 
The most and second most hydrogen-rich compounds created in this study are \ce{CH_{4}(H_{2})_{15}} with $\ce{CH_{4}}:\ce{H_{2}} = 1:15$ ($x = 0.9714$) and \ce{CH_{4}(H_{2})_{14}} with $\ce{CH_{4}}:\ce{H_{2}} = 1:14$ ($x = 0.9697$), and they show $\Delta H$ of 1.32\,mRy/atom and 0.78\,mRy/atom, respectively. 
We investigated the band gaps of 15 hydrocarbons (See Table \ref{results}). The band gap increases from 4.3 to 7.7\,eV with increase of $x$ from 0 to 0.75 and varies in the range of 7.9-8.1\,eV for the molecular compounds of \ce{C_{2}H_{6}} and \ce{CH_{4}} and in the range of 7.9-8.4\,eV for the molecular compounds of \ce{CH_{4}} and \ce{H_{2}}.

\section{Conclusion}
We proposed the evolutionary construction technique of the formation energy convex hull to search for thermodynamically stable phases in compounds. 
The potential candidates for the stable compounds are created by applying the three evolutionary operations, 
``mating'', ``mutation'' (permutation, distortion, reflection, modulation, addition, elimination, and substitution), and ``adaptive mutation'', 
to two compounds on and near the convex hull. 
In other words, the compositions and the structures for the next generation are created using the structural information of the stable and metastable compounds at the present generation. 
It is important for search algorithms to achieve a balance between exploration (global search) and exploitation (local search). In our search technique, 
the elimination and the substitution included in the mutation achieve the exploration with respect to the variation of the compositions, and the others correspond to the exploitation. In addition, the mating and the adaptive mutation increase the diversity with respect to the compositions and the structures. 
We emphasize that this technique can be applied to not only binary but also ternary or multinary system. 
Evolutionary algorithm usually requires setting many parameters such as 
heredity rate, permutation rate, mutation rate, population size, \textit{etc.}, to perform the structure search and tuning them to find the structures efficiently. 
Our technique reduces the burden on the parameter setting and tuning because the candidates are systematically created based on the rules shown in Figs. \ref{Fig-mutation} and \ref{Fig-nstep2}.  

We applied the evolutionary construction technique to C-H binary system at 10\,GPa and searched for thermodynamically stable compounds. As the results, in addition to diamond and $P2_{1}/c$ \ce{H_{2}}, we obtained 15 hydrocarbons with the convex hull distance less than 0.5\,mRy/atom: 
$(\ce{C_{2}H_{2}})_{n}$, $(\ce{C_{4}H_{6}})_{n}$, $(\ce{C_{2}H_{4}})_{n}$, \ce{C_{4}H_{10}}, \ce{C_{2}H_{6}}, \ce{C_{2}H_{6}CH_{4}}, \ce{C_{2}H_{6}(CH_{4})_{2}}, \ce{C_{2}H_{6}(CH_{4})_{6}}, \ce{CH_{4}}, \ce{(CH_{4})_{8}H_{2}}, \ce{(CH_{4})_{2}H_{2}}, \ce{CH_{4}H_{2}}, \ce{(CH_{4})_{5}(H_{2})_{6}}, \ce{(CH_{4})_{2}(H_{2})_{3}}, and \ce{CH_{4}(H_{2})_{2}}. These results indicate that 
hydrocarbons with $x \leq 0.6667$ are polymerized, those with $0.75 < x < 0.8$ form the molecular compounds of \ce{C_{2}H_{6}} and \ce{CH_{4}}, and those with $0.8 < x < 1$ form the molecular compounds of \ce{CH_{4}} and \ce{H_{2}}. 
Our calculations show that \ce{CH_{4}(H_{2})_{2}} ($x = 0.8889$) is the most hydrogen-rich compound on the convex hull at 10\,GPa, which is qualitatively consistent with the previous experimental and first-principles results. The compounds with $x > 0.8889$ are thermodynamically unstable and the hull distance increases with the increase of $x$. 
These results suggest that the evolutionary construction technique is useful to obtain the knowledge on thermodynamically stable compositions and structures. 

\begin{acknowledgments}
This study was supported by JSPS KAKENHI Scientific Research (C) (Grant No. 17K05541) and Scientific Research (S) (No. 16H06345), and gExploratory Challenge on Post-K Computerh (Frontiers of Basic Science: Challenging the Limits) and 
the Elements Strategy Initiative Center for Magnetic Materials (ESICMM) (No. JPMXP0112101004), through the Ministry of Education, Culture, Sports, Science and Technology (MEXT).
\end{acknowledgments}
\bibliography{References}

\end{document}